\documentclass[%
 aip,
 amsmath,amssymb,
 reprint,%
]{revtex4-1}

\usepackage{graphicx}
\usepackage{dcolumn}
\usepackage{bm}

\usepackage[utf8]{inputenc}
\usepackage[T1]{fontenc}
\usepackage{mathptmx}
\usepackage{etoolbox}
\usepackage{xcolor}
\usepackage{amsmath}
\makeatletter
\def\@email#1#2{%
 \endgroup
 \patchcmd{\titleblock@produce}
  {\frontmatter@RRAPformat}
  {\frontmatter@RRAPformat{\produce@RRAP{*#1\href{mailto:#2}{#2}}}\frontmatter@RRAPformat}
  {}{}
}%
\makeatother
\begin{document}

\preprint{AIP/123-QED}

\title{Electric Field Sensing via Rydberg Electromagnetically Induced Transparency Using Zeeman and Stark Effects}
\author{Yu-Chi Chen}
\author{Shao-Cheng Fang}
\author{Hsuan-Jui Su}
\author{Yi-Hsin Chen}
\email{yihsin.chen@mail.nsysu.edu.tw}
\affiliation{%
Department of Physics, National Sun Yat-Sen University, Kaohsiung 80424, Taiwan
 }%

\begin{abstract}
Rydberg-assisted atomic electrometry with thermal vapors offers a promising approach for detecting external electric fields. However, this technique presents significant challenges for measuring low frequencies due to the effects of metal-alkali atoms adsorbed on the interior surface of the vacuum chamber. In this work, we apply high-contrast Rydberg electromagnetically induced transparency (EIT) spectroscopy to systematically investigate these effects, including the influence of laser power and electric field strength. We demonstrate the ability to measure electric field frequencies ranging from 10 Hz to 1 MHz. Additionally, this study identifies a fundamental limit for data capacity in such measurements.
Furthermore, we propose a method for precise Stark shift measurements by locking the coupling laser to Zeeman-split Rydberg EIT peaks. Using the Zeeman shift in a reference cell as a stable frequency reference, we track Stark-induced shifts in a science cell and confirm excellent agreement with theoretical predictions. These results provide valuable insights for future precision measurement techniques and field sensing applications based on Rydberg atom systems.
\end{abstract}
\maketitle

\maketitle
\newcommand{\nonsupcite}[1]{\textnormal{\setcitestyle{numbers,square,comma,sort&compress}\cite{#1}}}

\newcommand{\Figsetup}{
\begin{figure}[b]
\centering
\includegraphics[width=1\columnwidth]{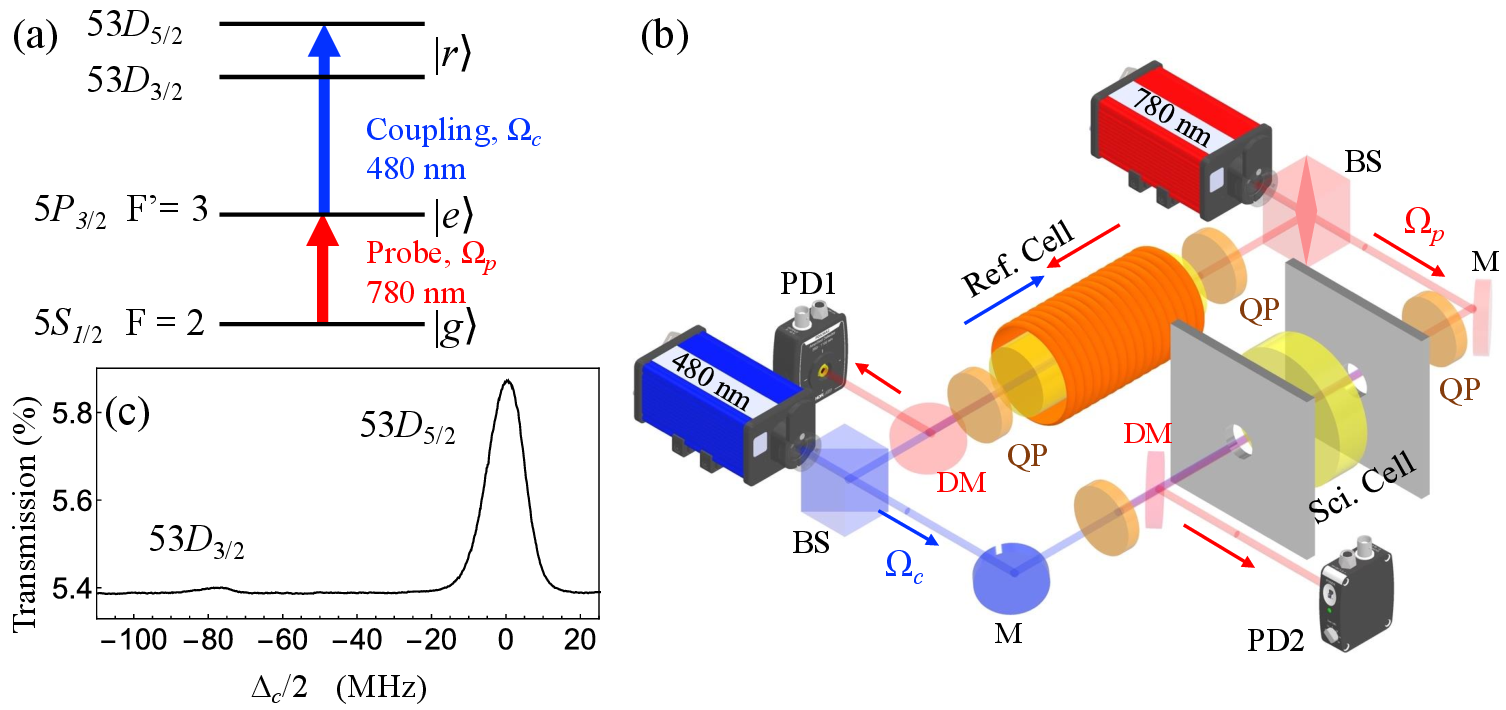}
\caption{(a) Energy levels of $\rm^{87}Rb$. The probe beam, with a wavelength of 780 nm, was frequency locked onto the transition from the ground state $|5S_{1/2}, F=2 \rangle$ to the excited state $|5P_{3/2}, F'=3 \rangle$. The coupling beam, with a wavelength of 480 nm, was swept to the transition from the excited state to the Rydberg states $|53D_{3/2}\rangle$ and $|53D_{5/2}\rangle$. The frequency difference between these two states is 78.3~MHz, used to calibrate the scanned frequency in spectra. (b) The schematic of the optical setup. The probe and coupling beams are counter-propagating, and their polarizations are both $\rm{\sigma^+}$. A reference cell with solenoid coils is used for frequency calibration. Two stainless steel electrodes are attached on each side of the science cell, with a distance of 20~mm between them. A heating block made of aluminum is placed outside the electrodes. DM: dichroic mirror; QP: quarter wave plate; BS: cubed beam splitter; PD: photodetector. (c) Rydberg EIT spectrum without external electric field was measured using probe field transmission in the reference cell.}
\label{fig:setup}
\end{figure}
}

\newcommand{\FigTwo}{
\begin{figure}[t]
\centering
\includegraphics[width=0.8\columnwidth]{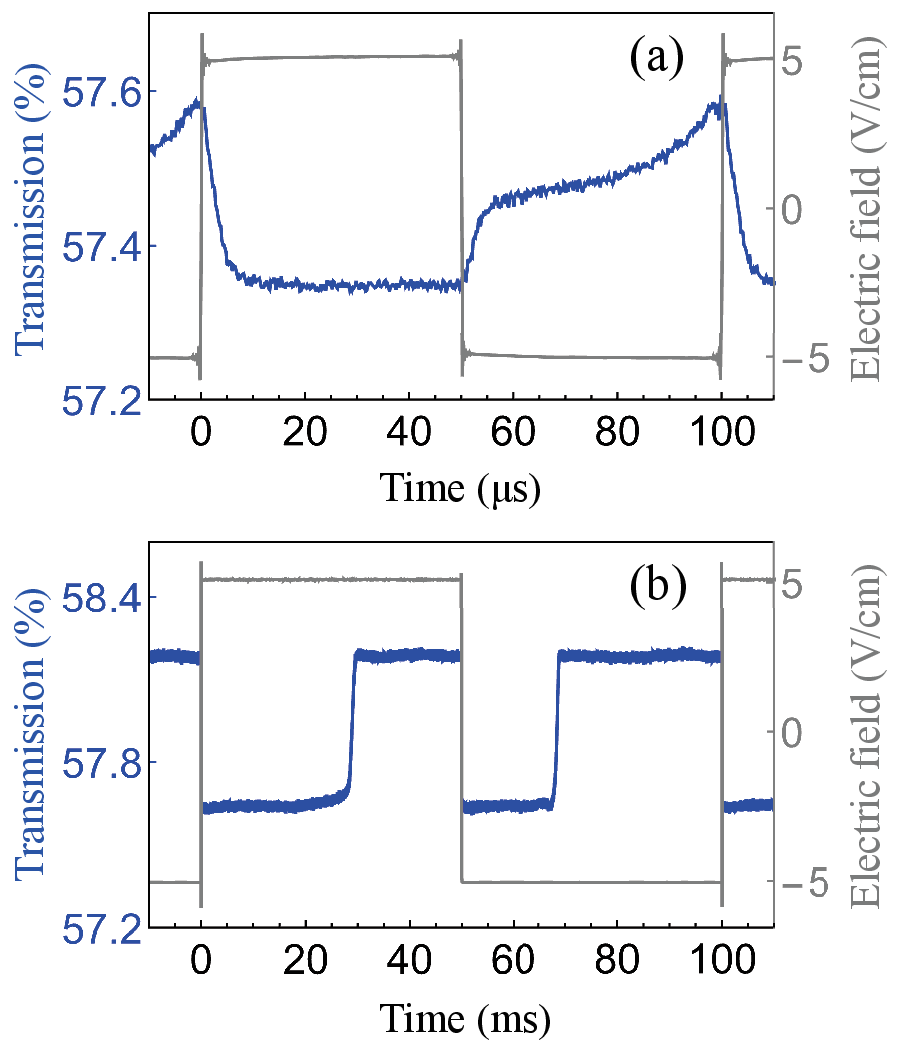}
\caption{Rydberg EIT peak signals varied under square waveform modulations of the electric field, with frequencies of 10 kHz in (a) and 10 Hz in (b). The electric field was generated by a function generator with 20~$\rm V_{pp}$, corresponding to the electric field of 10~V/cm. 
For fast modulation frequencies, we derive the transient time of $2~\mu$s from Eq.~(\ref{eq:exponential}), representing the duration during which the internal electric field is influenced by the external field. For slow modulation frequencies, the recovery times for the positive and negative electric fields are determined to be 30 ms and 20 ms, respectively.
The cell temperature was set to $65^{\circ}$C. }
\label{fig:Transient_time}
\end{figure}
}

\newcommand{\Figbeat}{
\begin{figure}[t]
\centering
\includegraphics[width=1\columnwidth]{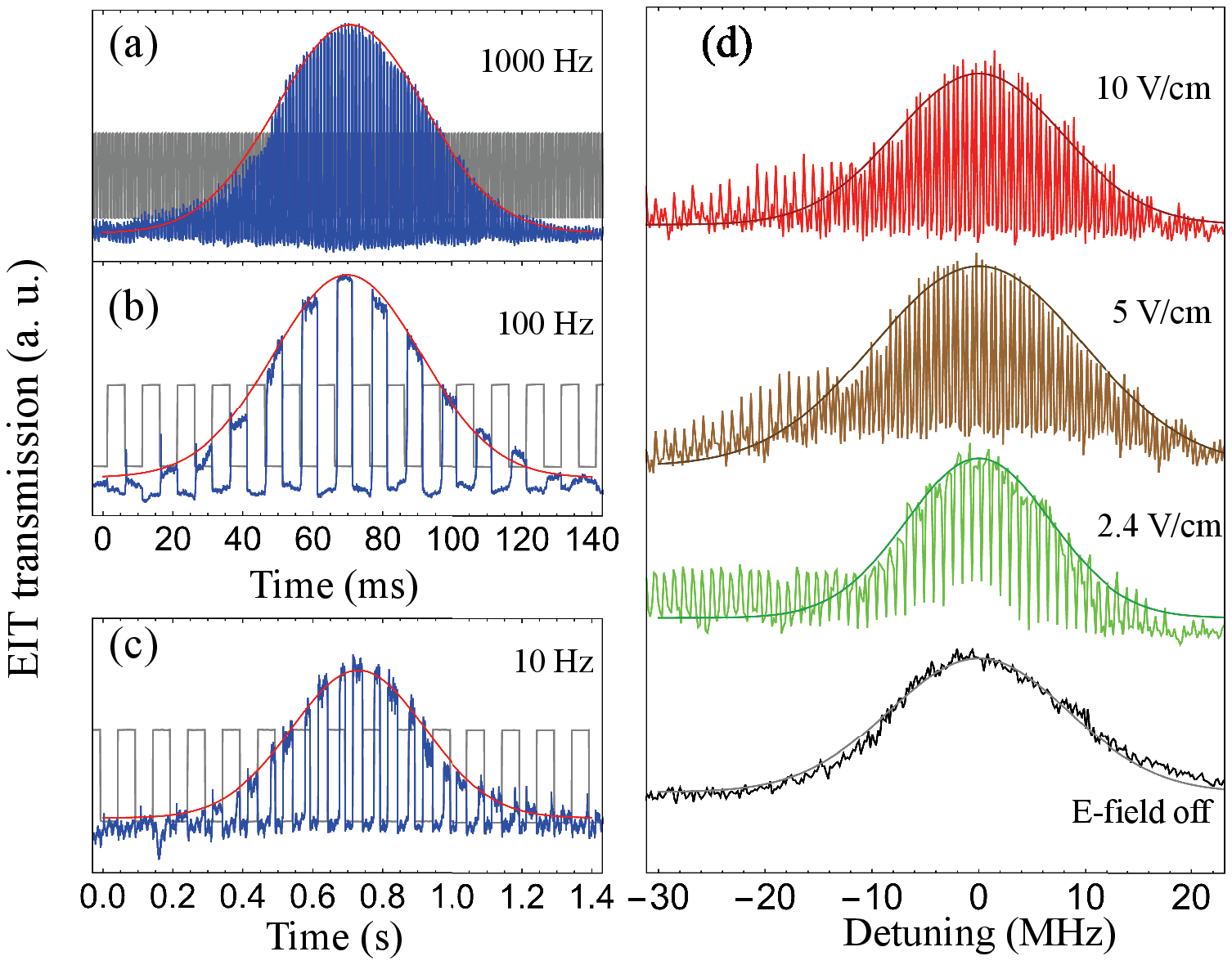}
    \caption{Rydberg EIT spectra under electric field modulations of 1 kHz (a), 100 Hz (b), and 10 Hz (c). The coupling laser frequency was scanned across the $|53D_{5/2}\rangle$ Rydberg state. Gray lines indicate the electric field signals. For sufficiently fast field modulations, e.g., above 100 Hz (Fig.~\ref{fig:beat}(b)), the EIT signals display a phase opposite to that of the electric field. For 10-Hz data, the coupling laser scanning time was increased by tenfold to improve clarity. (d) Rydberg EIT spectra with 1~kHz field modulation at field strengths of 10, 5, 2.4, 0~V/cm  from top to bottom. }
\label{fig:beat}
\end{figure}
}

\newcommand{\Figrecovery}{
\begin{figure}[t]
\centering
\includegraphics[width=1\columnwidth]{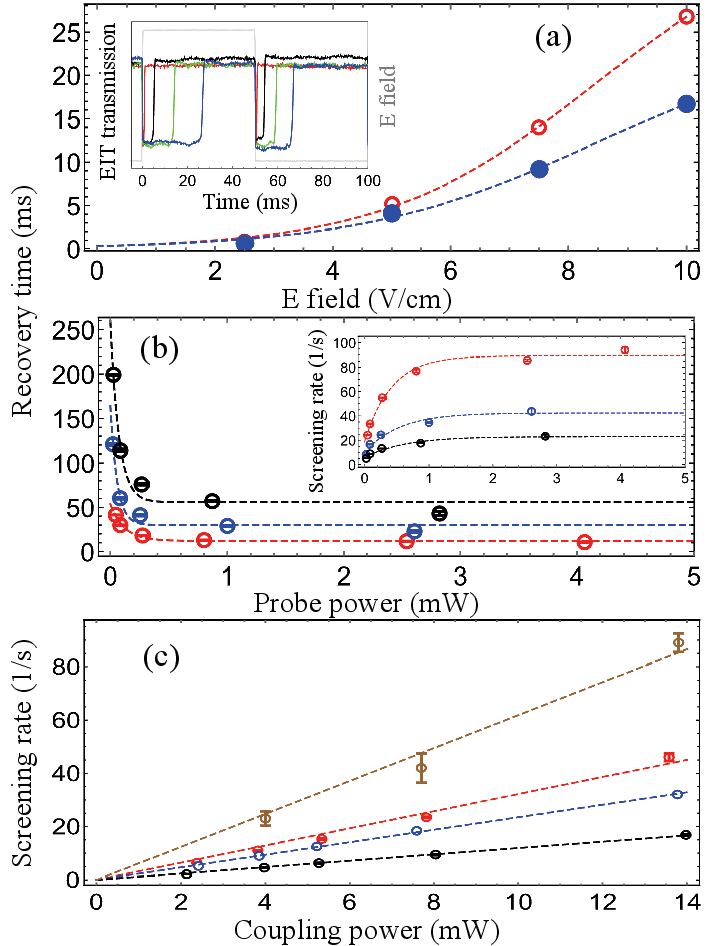}
\caption{The recovery times with varied electric field strength in (a) and varied probe power in (b). The screening rate (reciprocal of the recovery time) with varied probe power in the inset of (b) and varied coupling power in (c). The red-open circles and blue-solid circles in (a) represent the recovery times when the electric field is turned on in the positive and negative directions, respectively. The dashed lines are fitted using a logistic growth function to guide the eye. In (a), the probe and coupling powers were 0.16~mW and 14~mW (the strongest coupling power we applied in this study). In (b), the coupling power was 14~mW (red circles), 7.7~mW (blue), and 4.0~mW (black); and in (c), the probe power was 280 $\mu$W (red), 86 $\mu$W (blue), and 27~$\mu$W (black). The brown data represent the saturation screening rate $\gamma_0$ from the inset of (b). The field strength was set as 10~V/cm in both (b) and (c).}
\label{fig: recovery}                           
\end{figure}
}

\newcommand{\Figdensityplot}{
\begin{figure}[t]
\centering
\includegraphics[width=1\columnwidth]{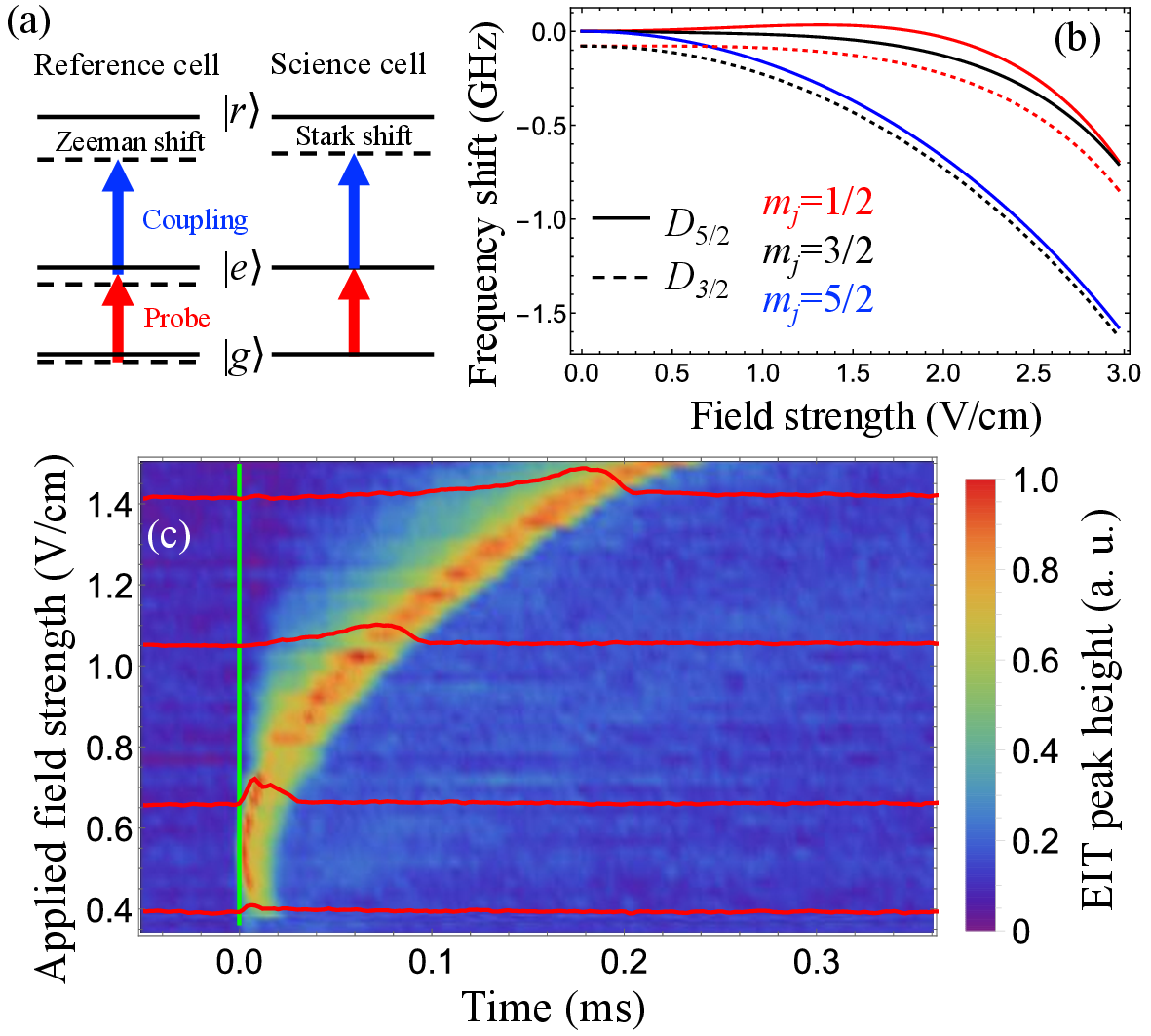}
\caption{(a) Schematic illustrating the energy level shifts caused by the Zeeman and Stark effects in both the reference and science cells. (b) Calculated Stark shifts of the $53D$ state as a function of electric field strength. (c) The normalized probe transmission signals for different electric field strengths, with the direction of the electric field was switched at $t=0$.}
\label{fig: densityplot}                           
\end{figure}
}

\section{Introduction}
Highly-excited Rydberg atoms exhibit a substantial electric dipole moment, leading to significant polarizability and robust interactions with light~\cite{gallagher,Vogt2006,saffman2010}. Moreover, utilizing the dipole-dipole interaction (DDI) between Rydberg atoms can tune the interactions among quantum bits, offering promising prospects for advanced quantum computing, including quantum gate operations~\cite{Jaksch2000,molmer2011}. The combination of Rydberg interactions and electromagnetically induced transparency (EIT) allows for the realization of photon quantum gates~\cite{Gorniaczyk2014,Tiarks2014,Tiarks2019}.
Furthermore, the interaction strength between Rydberg atoms can be enhanced through F\"{o}rster resonance, achieved by applying external fields based on the Stark effect~\cite{Ryabtsev2010,Ravets2014,Ravets2015,Wu2023,Sevincli2014,Tretyakov2014,Sylvain2017} with static fields~\cite{Ryabtsev2010,Ravets2014,Ravets2015,Wu2023}, microwaves~\cite{Tretyakov2014,Sevincli2014}, or optical fields~\cite{Sylvain2017}. 
\Figsetup

Rydberg atomic electrometry provides a method for measuring electric-field-induced energy shifts created by the Stark effect. Enhancing the sensitivity of this method is achievable through the use of the EIT phenomenon, facilitating optical detection of atomic interactions~\cite{Mohapatra2007,Jau2020,Ma2022,Li2023}. 
However, the presence of alkali-metal atoms on the inner surface of a vapor cell, due to a few layers of coating, induces non-zero conductivity on the glass~\cite{YuChi2022}.
The adsorption of alkali-metal atoms induces the formation of surface electric fields, which subsequently triggers an electric-field-screening effect due to laser irradiation absorption at the cell's surface~\cite{Kevin2018,Barredo2013}. The phenomenon, originating from the induced internal electric field by ions, also occurred with cold atoms in high vacuum chambers~\cite{McGuirk2004,Abel2011,Viteau2011}.
Several strategies are available to counteract the induced electric field within the cell: (1) utilizing externally connectable ring electrodes to apply electric fields within the cell~\cite{Ma2022}, (2) inserting film electrodes on the cell's interior~\cite{Barredo2013,Daschner2014}, (3) rapidly switching the electric field~\cite{Mohapatra2007}, (4) directly applying UV light, an Argon Laser, or heat to desorb surface metals~\cite{Alexandrov2002,Klempt2006,Sedlacek2016,Andersonw2018}, and (5) performing three-photon excitation with infrared laser fields~\cite{Lim2023}. The screening effect does not inhibit detection of high-frequency oscillating electric field signals ranging from MHz to THz. Notably, the technique can extend the detection spectrum of atomic electrometry, encompassing superlow-frequency (SLF) and extremely-low-frequency (ELF) bands~\cite{Li2023}. 

The detecting frequency of the atomic electrometry is limited by the time it takes for the induced electric field to respond to the external field. In principle, glass vapor cells are excellent electric insulators; however, the adsorption of alkali-metal atoms leads to nonzero conductivity. We define the recovery time as the time that free charges redistribute the glass surface to maintain equal potential on the conductive surface when an external electric field is applied. 
The recovery time was 500~$\mu$s in Ref.\nonsupcite{Kevin2018} and it can be varied from 100~$\mu$s to 700~$\mu$s, depending on the coupling field intensity in Ref.\nonsupcite{Mohapatra2007}. A longer recovery time of 200~ms is reported in Ref.\nonsupcite{Jau2020}.  
Our study utilizes Rydberg EIT resonances to characterize the cell's screening effect and charge dynamics as we vary parameters like electric field modulation frequency, field amplitude, and laser power.
The recovery time varied from 0.7 ms to 450 ms and a low frequency below 10 Hz was achieved. The results demonstrate significant progress in enhancing the performance of Rydberg atomic electrometry, offering valuable insights for future precision measurements and field sensing applications. 

Beyond dynamic field response measurements, we further explore the control of static energy level shifts using the combined Zeeman-Stark effect in Rydberg EIT spectra. To compensate for internal screening fields from surface-adsorbed charges, we employ a frequency-locking technique that uses Zeeman-split Rydberg EIT peaks in a reference cell as a stable frequency reference. This allows us to track Stark shifts in a science cell where external electric fields are applied. Through this method, we can determine both the actual Stark shift and the characteristic recovery time of the internal field following electric field switching. The results agree well with theoretical predictions and reveal how screening dynamics influence the operation time period for observing controlled interactions, such as F\"{o}rster resonances. This capability improves the precision and adaptability of Rydberg-based field sensing in environments.

\section{\label{sec:level1}Experimental Setup}

We carried out the electric field measurement in a Rb vapor cell. The energy levels of $\rm^{87}Rb$ involving Rydberg state and laser excitations are shown in Fig.~\ref{fig:setup}. 
The probe beam, with a wavelength of 780 nm, was locked onto the transition from the ground state $|5S_{1/2}, F=2 \rangle$ to the excited state $|5P_{3/2}, F'=3 \rangle$ via the saturation absorption spectrum, which is not shown in the figure. Meanwhile, the coupling beam, with a wavelength of 480 nm, was either frequency scanned across or locked to the transition of the excited state to the Rydberg state $|53D_{5/2}\rangle$. The probe and coupling beams were counter-propagated through a reference cell and a science cell, and their polarizations were both $\rm{\sigma^+}$ to achieve the maximum Rydberg EIT signal~\cite{Su2022, Su2022_2}.
These beams were collimated to a full width at $e^{-2}$ maximum of 0.81~mm. 
To control the sheet resistance on the inner surface of the glass cell, we varied the coupling laser power from 2.4 mW to 14 mW and changed the probe laser power from 0.023 mW to 4.1 mW. The optimized probe power was 0.088 mW to achieve the highest EIT peak height~\cite{Su2022}.
The Rydberg EIT spectrum in Fig.~\ref{fig:setup}(c) was obtained by measuring the transmission of the probe field in the absence of electric and magnetic fields in the reference cell. The scanned detuning was calibrated based on the frequency difference between $|53D_{3/2}\rangle$ and $|53D_{5/2}\rangle$~\cite{Mach2011}. Using the EIT peak in the reference cell, we can compare the frequency shift caused by the external electric field. 

We constructed the Rb science cell with flat windows of 25~mm in diameter. The thickness of the cell's windows is 3~mm, while the overall length of the cell is 19.2~mm (with a path length of 13.2~mm). The vacuum pressure was 2.2$\times10^{-7}$ torr before sealing the cell. The cell was made of borosilicate glass and filled with a natural abundance of Rb vapors. Two stainless steel electrodes were attached to each side of the cell. Both electrodes' dimensions are $50\times50~\rm mm^2$ to generate a uniform electric field in the laser propagation direction. Laser beams are transmitted through electrode holes of 4 mm in diameter, with the electric field aligned parallel to the laser beams.
An aluminum heating block is positioned adjacent to the electrodes.
Optical density (OD) is a factor that increases the sensitivity of the EIT and the electrometry. It can be controlled by vapor temperature and determined by the probe field absorption signal at the resonance frequency. The cell temperature is set at 65$^\circ$C or 75$^\circ$C. 

\section{Results} \label{sec:results}
A scanned coupling field, combined with a DC electric field, was employed to investigate the Stark shift in the Rydberg EIT spectrum. Theoretical calculations using the Alkali Rydberg Calculator (ARC) Python package~\cite{SIBALIC2017} predict a Stark shift of a few hundred MHz for $|m_j|=5/2$ under an electric field of 1 V/cm. Details can be found in Fig.~\ref{fig: densityplot}(b). While the experimental results did not reveal a significant shift in the spectrum, this unexpected outcome provides valuable understanding into the system dynamics. It suggests the possible influence of a screening effect, potentially caused by Rb coatings excited by the coupling field at a wavelength of 480~nm. This excitation generates an internal electric field that shields the external field, revealing a complex interplay with surface effects. These findings highlight the importance of surface interactions and open opportunities to further explore screening phenomena in Rydberg systems.

In order to gain insights into the screening effect and charge dynamics within the cell, we locked the coupling field to the peak transmission of Rydberg EIT and employed a square waveform to switch the electric field, as illustrated by the black line in Fig.~\ref{fig:Transient_time}. Subsequently, we detect the transmission signal from the probe field at resonant EIT, represented by the blue line. 
When a sufficiently high-frequency electric field is applied, e.g., 10 kHz in Fig.~\ref{fig:Transient_time}(a), the EIT transmission does not have enough time to respond adequately to the rapid switching of the electric field. As a result, the transmission experiences a sudden drop when the electric field is turned on and a sudden increase when the electric field is turned off, exhibiting an opposite response to the field's state.
Here we introduce the transient time, denoted as $\tau$, which characterizes the duration for the EIT signal $T$ decay to $e^{-1}$ of its amplitude $A$, as expressed by an exponential function 
\begin{equation}
           T(t)=Ae^{-\frac{t}{\tau}}+y_0. 
 \label{eq:exponential}
\end{equation}
The probe field offset transmission, $y_0$, indicates that the Stark shift occurs in the steady state.
The derived transient time is approximately a few $\rm{\mu}$s.
When the switching frequency of the electric field exceeds the reciprocal of the transient time, the internal electric field is unable to respond effectively. It constrains the upper boundary of the frequency within the range of MHz by this technique. 

The vapor cell is made from a glass material that is an excellent insulator. It can have non-zero conductivity due to the optically induced adsorption of alkali-metal atoms on its interior surfaces. 
If the electric field is applied at a low frequency, e.g., 10 Hz, the EIT transmission suddenly jumps as shown in Fig.~\ref{fig:Transient_time}(b). Following each jump, the EIT transmission took several tens of milliseconds to return to its original level.
This indicates that the free charges on the inner surface redistribute over a certain timescale, defined as the recovery time, leading to an equal electric potential on the surface and shielding of the external electric field. Take the measurement in Fig.~\ref{fig:Transient_time}(b) as an example. The recovery time is longer when the electric field is positive (30~ms) compared to the negative one (20~ms). Detailed discussion about the effect of the recovery time can be found in Sec.~\ref{sec:discussion}. 

\FigTwo

\Figbeat     
To ensure that the electric field induces a Stark shift and influences the Rydberg EIT, we simultaneously switched the electric field while sweeping the coupling laser across $|53D_{5/2}\rangle$ Rydberg state. 
Figures \ref{fig:beat}(a), \ref{fig:beat}(b), and \ref{fig:beat}(c) are the spectra with the electric field modulation frequencies of 1 kHz, 100 Hz, and 10 Hz. For 10-Hz data (in \ref{fig:beat}(c)), the coupling laser scanning time was increased tenfold to improve clarity. 
In the presence of an external electric field, the Rydberg energy shifts from resonance, resulting in a two-level absorption, subsequently causing a reduction in the EIT signal. 
For fast modulation, e.g., above 100 Hz, the induced energy shift cannot return to its original state in time, causing the EIT signals to display a phase opposite to that of the electric field signal. As a result, the Rydberg EIT spectrum shows a square-wave modulation. When the duty cycle of the EIT signal is approximately $50\%$, as in the case of a 10 Hz field modulation, it sets a lower limit on the achievable frequency.
Figure \ref{fig:beat}(d) presents the Rydberg EIT spectra under a 1 kHz electric-field modulation, corresponding to different field strengths. 
The solid lines represent Gaussian fits to the EIT resonance envelopes. 
At the intensity of 10 V/cm, the linewidth broadened due to the non-uniform distribution of charges within the cell, resulting in an inhomogeneous electric field.
In addition to the square modulation observed on the Rydberg resonance, the Stark effect is evident in the detuning range of 0 to -30 MHz, where long-tail EIT signals appear above the Gaussian fitting lines. The EIT resonance shifts in the same direction as the electric field due to the quadratic dependence of the Stark shift on the field strength, expressed as $\delta\propto E^2$. This quadratic relationship arises from the second-order perturbation of energy levels in the Stark effect, making the resonance shift proportional to the square of the electric field and symmetric with its magnitude.

\Figrecovery
\section{Discussion} \label{sec:discussion}
The lower limit of measurable field frequency is determined by the ability of suppressing external electric fields. A square wave applied to the external field provides the measurement of the response time required for the internal electric field to develop. This makes it an effective method for characterizing low-frequency operation. 
However, the detection of higher frequencies is limited by the transient time, as described in Eq.~(\ref{eq:exponential}). This study shows that transient times range from 1 to 10~$\mu$s, with minor fluctuations attributed to vapor temperature and major variations correlated with EIT peak signals. These variations set an upper limit of 1 MHz for the detectable frequency. Consequently, our study establishes a fundamental limit for data capacity and experimentally demonstrates its validity across bandwidths from 10 Hz to 1 MHz.

The shielding recovery time refers to the period required for free charges to redistribute on the glass surface, ensuring an equipotential condition on the conductive surface when an external field is applied. To further investigate the factors influencing the recovery time, we first vary the field strength while keeping the probe and coupling powers at 0.16 and 14~mW. The extracted recovery times are shown in Fig.~\ref{fig: recovery}(a).
The growth behavior suggests that electron concentration increases with field strength, leading to an interplay between carrier generation, mobility, and interactions in response to external fields. Mobility refers to the ability of electrons to move across the glass surface under an applied field. A stronger field can enhance charge carrier mobility, facilitating more efficient transport and accumulation.
Applying a strong electric field prolongs the recovery time due to significant energy level shifts and disturbances in the internal field. In Fig.~\ref{fig: recovery}(a), the red-open and blue-solid circles represent the recovery times at the positive and negative field transitions, respectively. Measurements show that the recovery time for the positive field transition is approximately 1.5 times that of the negative one, likely due to an inhomogeneous field generated by the electrodes. This effect was confirmed by reversing the field direction in space.
Thus, in the following discussion, we focus only on the analysis for the positive electric field direction. Physical limitations such as finite charge carrier availability, recombination, surface space-charge effects, and reduced charge mobility at high electron densities lead to saturation effects that ultimately limit the recovery time, despite increasing laser power or electric field strength.

We then adjust the probe and coupling powers to measure the recovery time under a field strength of 10 V/cm, as shown in Figs.~\ref{fig: recovery}(b) and \ref{fig: recovery}(c). As the probe power increases, more photons interact with free electrons, accelerating their equilibration and resulting in a shorter recovery time.
Here, we define the screening rate of the electric field as the reciprocal of the recovery time. In Fig.~\ref{fig: recovery}(b), the coupling powers were 14~mW (red circles), 7.7~mW (blue), and 4.0~mW (black). 
The screening rate grows exponentially with probe power and eventually saturates. At low probe power, the screening rate increases because additional photons contribute to the excitation of free electrons, enhancing charge redistribution. However, as probe power continues to rise, the number of available electrons becomes limited, and recombination processes start to counteract further excitation, leading to saturation.
Dashed lines indicate the best fit using the function $A\cdot$ Exp[-$P_{p}$/$\tau_p$]+$\gamma_0$, where $A$ is a constant, and $P_{p}$ is the probe power. The saturation screening rates, denoted as $\gamma_0$, are 89, 42, and 23 (1/s), with corresponding $\tau_p$ values of 0.37, 0.54, and 0.60 mW for the strongest to the weakest coupling power. The saturation screening rate, reflecting the current in the photoelectric effect, is predominantly influenced by the coupling laser power (480~nm). $\tau_p$ serves as a parameter representing the characteristic constant associated with the photoelectric current's response to the probe laser power. A smaller $\tau_p$ implies a faster response of the photoelectric current to change in laser power. Therefore, stronger laser powers result in faster screening rates, corresponding to shorter recovery times.

Next, we investigate the relationship between the screening rate and coupling power while keeping the probe power constant. In Fig.~\ref{fig: recovery}(c), the screening rate demonstrates a linear increase with the coupling power, maintaining probe powers of 280 $\mu$W (red circles), 86 $\mu$W (blue), and 27 $\mu$W (black). The brown data represent the saturation screening rate $\gamma_0$ from the inset of Fig.~\ref{fig: recovery}(b). The corresponding slopes are 6.2, 3.2, 2.4, and 1.2 (1/s/mW) from top to bottom. 
This observation confirms that the screening rate of excited Rb atoms is influenced by the 780~nm probe laser and the 480~nm coupling laser.
A previous remark suggested that a 780~nm photon alone may not directly induce electronic transitions in solid-state Rb metal. However, its absorption may create localized effects, transferring energy to electrons and increasing their kinetic energy. To test this, we introduce an additional laser to excite $^{85}$Rb atoms, generating more free electrons. This led to a shorter recovery time, confirming that additional excitation enhances charge generation and influences the screening process. When exposed to 480 nm photons, these energized electrons escape more easily from the surface, further enhancing the screening effect. By adjusting the laser powers, the recovery time can be tuned from 1 ms to 400 ms. 
Consequently, the detectable frequency range of Rydberg atomic electrometry extends into the extremely low-frequency (ELF) band, highlighting the role of photon-induced charge dynamics. Beyond the photoelectric effect, photon interactions can also influence the adsorption and desorption of atoms on the surface. In particular, desorbed atoms may further modify the charges by altering surface conditions and redistributing free carriers. This leads to an additional mechanism affecting the screening process. There is also a phenomenon known as light-induced atom desorption (LIAD), which is commonly used to flexibly control atomic density~\cite{Alexandrov2002,Klempt2006}. 

\Figdensityplot
Finally, we derive the magnitude of the Stark shift by frequency-locking the 480~nm coupling laser to the Zeeman sublevel transitions between $|5P_{3/2}\rangle$ and $|53D_{5/2}\rangle$ states. In the presence of a magnetic field generated by solenoid coils, the EIT spectrum splits with the separation proportional to the magnetic field strength. Further details of this technique are provided in our previous study~\cite{Su2022}.
In this configuration, the laser was locked to the red-detuned Zeeman-split EIT peaks. A recent study demonstrated a similar laser locking technique using Zeeman-split Rydberg EIT peaks, achieving a wide continuous tuning range of $0.6$ GHz along with enhanced lock stability and reduced linewidth compared to conventional methods~\cite{Sile2024}. 
In our setup, a magnetic field of 71 Gauss was applied in the reference cell, producing a Zeeman shift of 260 MHz for 
$m_j=-5/2$, which served as the frequency reference for tracking the Stark-shifted EIT peaks. An additional electric field was then applied in the science cell to induce a Stark shift.
The schematic in Fig.~\ref{fig: densityplot}(a) illustrates the energy level shifts produced by the Zeeman and Stark effects in both the reference and science cells. 
Figure \ref{fig: densityplot}(b) presents the calculated Stark shifts for the $|53D_{m_j}\rangle$ states, using the ARC Python package~\cite{SIBALIC2017}. 

To investigate the time-dependent response of the Stark shift, Fig.~\ref{fig: densityplot}(c) presents the probe transmission signal under different electric field strengths. At $t=0$, the direction of the applied electric field was inverted. During the recovery time, the external field remains, while the internal field has not yet fully established. This imbalance shifts the system into two-photon resonance, producing the EIT peak seen in the bottom two red traces in Fig.~\ref{fig: densityplot}(c).
If the electric field strength is sufficiently large, as in the top two traces, no probe signal is observed immediately after $t=0$. Once the internal field rapidly builds up and restores the energy levels to their original ones, the EIT peak appears briefly when two-photon resonance condition is satisfied. 
Since the direction of the electric field was inverted at $t=0$, the effective change in field strength experienced by the vapor is twice the applied value. For example, when the applied field was 0.65 V/cm, corresponding to an effective change of 1.3 V/cm, a double-peak structure appeared.
The relative Stark shift at 1.3 V/cm for $m_j=5/2$ is approximately $-280$ MHz, providing clear evidence that the Stark effect causes the splitting of $m_j=5/2$ sublevels. As the electric field strength increases further, the width of the EIT signal broadens compared to the case without an applied field, consistent with observations reported in Ref.\nonsupcite{Holloway2022}. As expected, we observed that the EIT peak shifts from its zero-field position at $-260$~MHz (due to the Zeeman shift) to around $-280$~MHz under the applied field (due to the Stark shift).
This approach demonstrates a reliable method for measuring Stark shifts by combining Zeeman-split laser locking with controlled electric field modulation. The observed frequency shifts agree well with both Zeeman and Stark predictions, confirming the accuracy and effectiveness of this technique for probing Rydberg energy level shifts.

To further highlight the application of this measurement technique, it is important to note that precise control and characterization of Stark shifts are essential for studying F\"{o}rster resonances in Rydberg atoms. An applied electric field is typically used to tune Rydberg energy levels into resonance; however, the resulting internal screening field tends to counteract the external field, complicating control of the interaction strength. The Zeeman–Stark effects study presented here demonstrates that by tracking the Stark-induced frequency shifts and the recovery dynamics of the internal field, it is possible to accurately determine both the actual resonance frequency and the effective time window before the internal field stabilizes. This provides critical information for optimizing the timing and conditions for observing F\"{o}rster interactions in Rydberg ensembles, particularly in environments where surface charges and screening effects are non-negligible.

\section{Conclusion and Outlook}

Our experiments in a Rb vapor cell show the ability of Rydberg EIT spectroscopy for probing electric field-induced effects in atomic systems. We successfully identified the screening effect induced by external electric fields and Rydberg excitation, clarifying its origin in the charge dynamics driven by the photoelectric effect of the Rb coating under the influence of the lasers.
We demonstrated the capability to measure frequency electric fields from 10 Hz to 1 MHz. The detection range of this atomic electrometry is fundamentally constrained by the recovery and transition times of the shielding process, which are strongly influenced by the laser powers as well as the applied electric field strength. Our systematic study revealed that these parameters can be used to tune the recovery time and screening rate, thereby extending the measurement range. Notably, this approach enables reliable detection of fields within the extremely-low-frequency (ELF) band, presenting a promising technique for electric field measurements outside vacuum environments.
In addition, we demonstrated a method for precise Stark shift measurements by locking the coupling laser to Zeeman-split Rydberg EIT peaks. Using a reference Zeeman shift as a stable frequency target, we tracked Stark shifts in the science cell and observed good agreement with theoretical predictions. 
This approach holds potential for applications in precision electrometry, RF field detection, and surface charge studies, particularly in environments where conventional sensors are limited. The wide detection range, tunability, and practical performance make it a promising platform for quantum sensing.

\begin{acknowledgments}
This work was supported by Grants Nos. 112-2112-M-110-008 and 112-2123-M-006-001 of the National Science and Technology Council, Taiwan. The authors thank Prof. Tsu-Chiang Yen for the valuable comments on the study.
\end{acknowledgments}
\section*{AUTHOR DECLARATIONS}
\subsection*{Conflict of Interest}
The authors have no conflicts to disclose.
\section*{Data Availability Statement} The data that supports the findings of this study are available within the article. 

\bibliography{Forsterreference.bib}
\end{document}